\documentclass{iau}
\usepackage{graphicx}

\title[Warm Debris Disks with WISE] 
{Sensitive Identification of Nearby Debris Disks via Precise Calibration of WISE Data}

\author[R. Patel, S. Metchev, A. Heinze \& J. Trollo]   
{Rahul Patel$^1$, Stanimir Metchev$^{1, 2, 3}$, Aren Heinze$^1$ \and Joe Trollo$^1$}

\affiliation{$^1$Stony Brook University - email: {\tt rahul.patel.1@stonybrook.edu}\\
$^2$Univeristy of Western Ontario - email: {\tt smetchev@uwo.edu}\\
$^3$Centre for Planetary Science and Exoploration}

\pubyear{2015}
\volume{314}  
\pagerange{119--126}
\jname{Young Stars \& Planets Near the Sun}
\editors{J. H. Kastner, B. Stelzer, \& S. A. Metchev, eds.}
\begin{document}

\maketitle

\begin{abstract}
Using data from the WISE All-Sky Survey, we have found $>100$ new infrared excess sources around main-sequence Hipparcos stars within 75~pc. Our empirical calibration of WISE photospheric colors and removal of non-trivial false-positive sources are responsible for the high confidence ($>$99.5\%) of detections, while our corrections to saturated W1 and W2 photometry have for the first time allowed us to search for new infrared excess sources around bright field stars in WISE. The careful calibration and filtering of the WISE data have allowed us to probe excess fluxes down to roughly 8\% of the photospheric emission at 22$\mu m$ around saturated stars in WISE. We expect that the increased sensitivity of our survey will not only aid in understanding the evolution of debris disks, but will also benefit future studies using WISE.


\end{abstract}

\firstsection
\section{Introduction}

    Micron sized dust grains in debris disks are generated from the grinding down of asteroids and comets, whose collisions are induced by the gravitational stirring of planetary bodies. Thus, the presence of a debris disk can act as a signpost for planetary systems. The majority of disks, discovered by IRAS, ISO, and the Spitzer Space Telescope (see reviews by \cite[Zuckerman 2001]{Zuckerman2001} \& \cite[Wyatt 2008]{Wyatt 2008}) are cold disks ($T_D<100$~K) analogous to the Kuiper Belt in the Solar System. 
Warm disks ($T_D\sim 150-300$~K) probe activity in the terrestrial planet zone of the star. These warm disks are typically identified by an IR excess between 10--30$\mu m$, but are relatively rare ($\sim$4\% for solar type stars; \cite[Trilling et al. 2008]{Trilling2008}). 
    
    To compensate for the low incidence of warm disks, data from the Wide Field Infrared Survey Explorer all-sky mission (WISE; \cite[Wright et al. 2010]{Wright2010}) can be used to identify a large number of warm disks. This is due to the mission's increased sensitivity and resolution compared to the Infrared Astronomical Satellite (IRAS). The contemporaneous coverage of WISE in the W1, W2, W3, and W4 bands (3.4, 4.6, 12, and 22$\mu m$, respectively) has previously been used to identify new disks using the W3 and W4 bands. In our survey, we include bright saturated stars in WISE as well as decrease our sensitivity to large sources of false-positives to probe faint dust levels, making this work complementary to previous studies. 
\vspace{-0.5 cm}
    
\section{Identification of Bona-Fide Excesses}

\begin{figure}
\begin{center}
 \begin{tabular}{cc}
 \includegraphics[width=0.5\textwidth]{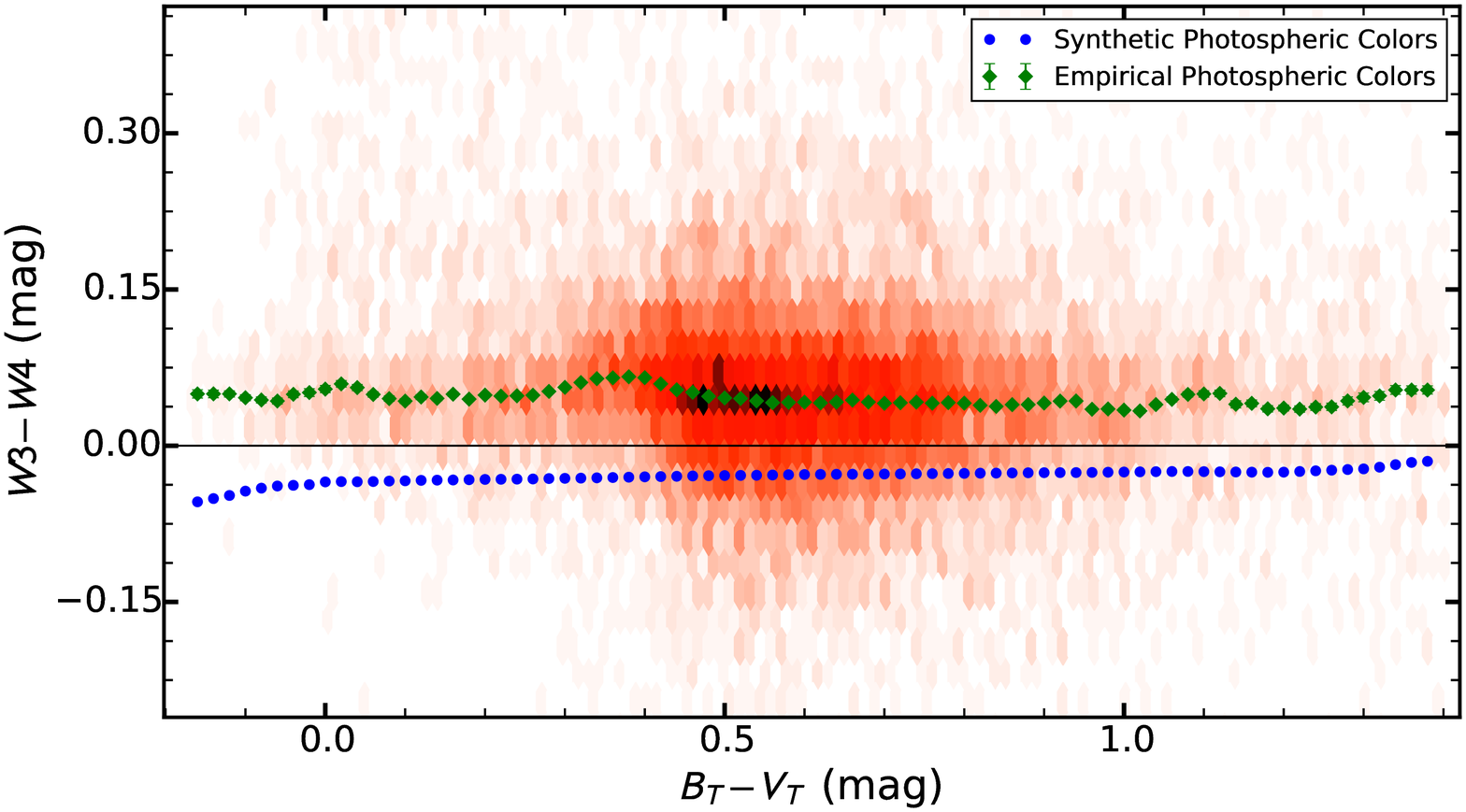}
 \includegraphics[width=0.5\textwidth]{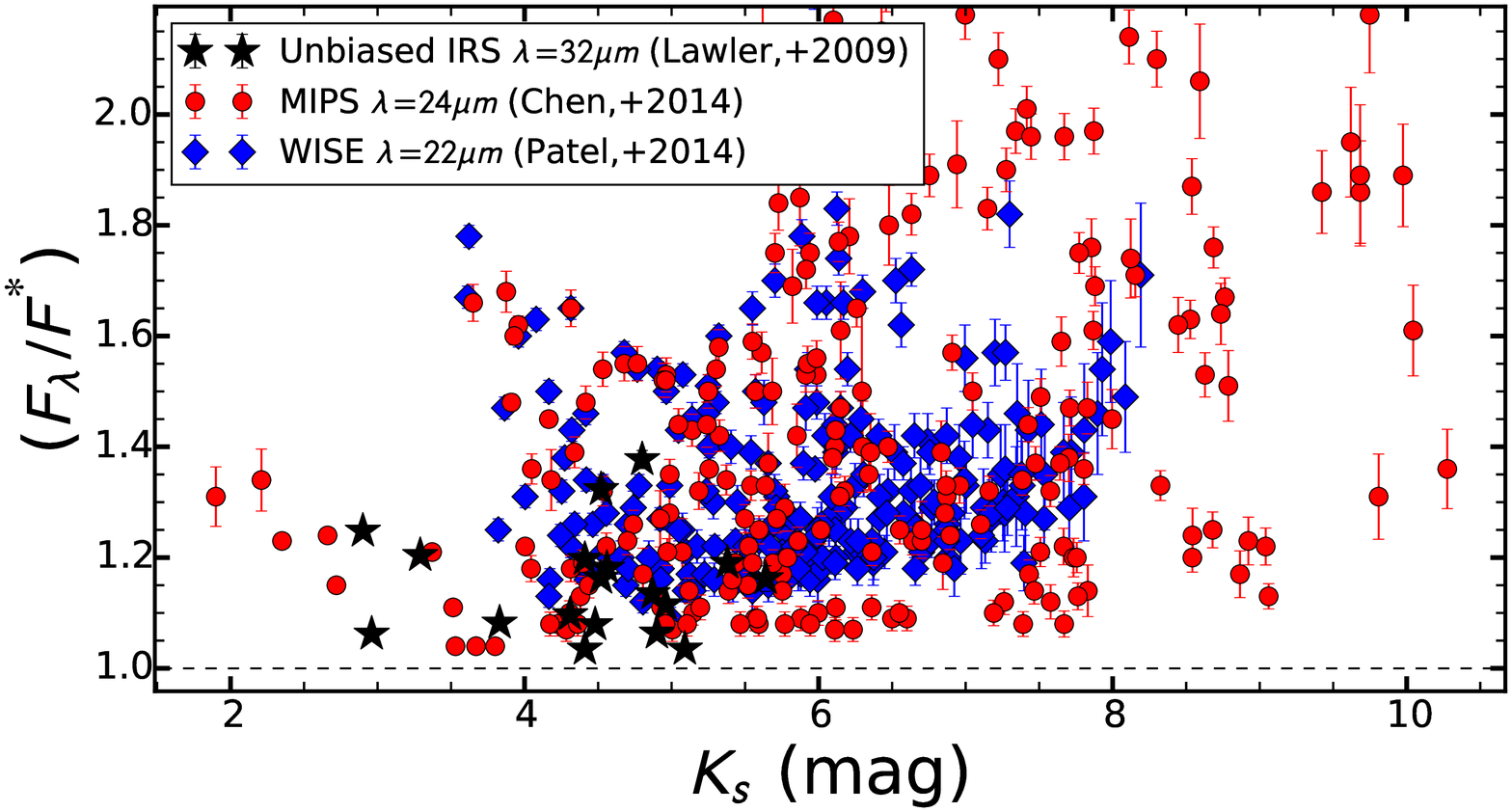}
\end{tabular}
\caption{\textbf{Left:} A color-color diagram for Hipparcos main-sequence stars within 120~pc, showing the trend between IR and $B_T-V_T$ photospheric colors. We subtracted this dependence to eliminate the photospheric contribution, and the residuals (excess) were used to assess the 99.5\% confidence level above which stars were selected as excess candidates. The synthetic colors were derived using NextGen models convolved to the WISE bandpasses. \textbf{Right:} Comparison of relative fluxes for excesses identified from Spitzer/MIPS (\cite[Chen et al. 2014]{Chen2014}), unbiased sample of Spitzer/IRS (\cite[Lawler et al. 2009]{Lawler2009}) and our own WISE detections.}

\end{center}
\label{fig1}
\end{figure}

    We used five WISE colors to search for excesses at W3 or W4 (W1-W4, W2-W4, W3-W4, W1-W3 and W2-W3). The details of our sample can be found in \cite[Patel et al. (2014)]{Patel2014}. In short, we used main-sequence Hipparcos stars within 120~pc, that lie outside the galactic plane ($|b|>5^{\circ}$). We included stars up to 4.5~mag in W1 and 2.7~mag in W2 by applying corrections to saturated photometry (Fig.~2 in \cite[Patel et al. 2014]{Patel2014}). Using the larger 120~pc set of stars as a parent sample,  we accurately determined the photospheric behavior of WISE colors. The left panel in Fig.\,\ref{fig1} shows the empirically identified photospheric trend we used to subtract from our WISE colors. Over-plotted are synthetic photospheric colors that show a systematic offset from the empirical trend. Without accounting for the difference between the empirical and synthetic photospheric colors, the magnitude of the WISE excess can be overestimated. 
   
    Candidate debris disk stars were selected from each color based on the significance of their color excess at confidence levels $>99.5$\%. Along with a large array of filters to remove false-positives (see Sec. 2 in \cite[Patel et al. 2014]{Patel2014}), we verified our excess detections by assessing the significance of their weighted color excess \cite[(Patel et al. 2015)]{Patel2015}. We also identified and removed contaminated stars by using higher resolution images from the unWISE survey \cite[(Lang 2014)]{Lang2014} and compared the relative centroid offsets between W3 and W4 for each star. Significant shifts indicated stars whose photometry were contaminated by nearby or background objects within the WISE beam.
\vspace{-0.5 cm}

\section{Results and Discussion}

    We identified 237 excesses at either W3 or W4 within 75~pc of the Sun. Our survey has increased the total number of disks within 75~pc by 25\%. Our study has also increased the census of excesses at $10-30\mu m$ by 35\%. We show that 16--22\% of A stars and 1.4--1.8\% of FGK stars possess W4 excesses. The incidence of W3 excesses is much smaller; 0.8--1.0\% of A stars and 0.04--0.08\% of solar type stars possess a W3 excess. We detect IR excesses that are on average 30\% above the photosphere at 22$\mu m$, and in a few cases as little as 8\% above the photosphere (right panel Fig.\,\ref{fig1}). Although WISE is not as sensitive to excesses as Spitzer/MIPS and Spitzer/IRS, this is not surprising, considering that those surveys were pointed and deeper.

\end{document}